\documentclass[pra,aps]{revtex4}

\usepackage{url}
\usepackage{amssymb,amsmath}
\usepackage{mathrsfs}
\usepackage{ifthen}
\usepackage{epsfig}

\newcommand{\be}{\begin{align}}
\newcommand{\ee}{\end{align}}
\newcommand{\mats}{\begin{bmatrix}}
\newcommand{\mate}{\end{bmatrix}}
\newcommand{\vecs}{\begin{bmatrix}}
\newcommand{\vece}{\end{bmatrix}}

\newcommand{\figs}{\ifthenelse{\equal{\showFigs}{true}}} 
\newcommand{\ignore}[1]{}

\newtheorem{theorem}{Theorem}[section]
\newtheorem{proposition}[theorem]{Proposition}

\newtheorem{lemma}[theorem]{Lemma}

\newcommand{\qedsymb}{\hfill{\rule{2mm}{2mm}}}
\newenvironment{proof}[1][]{\begin{trivlist}
\item[\hspace{\labelsep}{\bf\noindent Proof#1:\/}] }{\qedsymb\end{trivlist}}

\begin{document}
\title{Expectations of two-level telegraph noise}
\author{Jesse Fern$^1$}
\affiliation{$^1$Department of Mathematics, University of California, Berkeley, California, 94720}

\date{\today}

\begin{abstract}
We find expectation values of functions of time integrated two-level telegraph noise. Expectation values of this noise are evaluated under simple control pulses. Both the Gaussian limit and $1/f$ noise are considered.  We apply the results to a specific superconducting quantum computing example, which illustrates the use of this technique for calculating error probabilities. 
\end{abstract}
\maketitle

\section{Introduction}

\begin{figure}[hbtp]
 \vspace{9pt}

  \centerline{\hbox{ \hspace{0.0in} 
   \epsfxsize=2.0in
    \epsffile{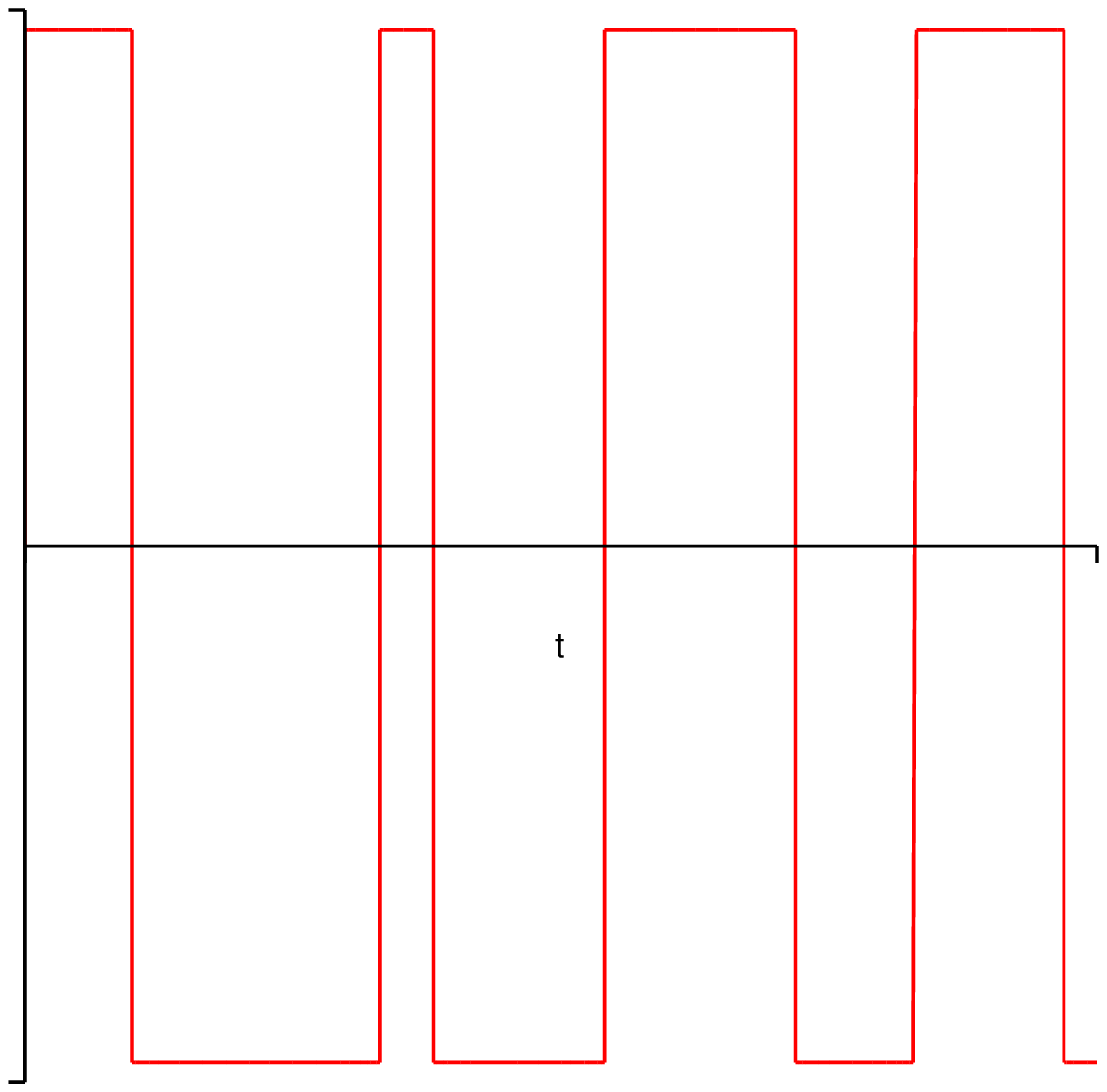}
    \hspace{0.25in}
    \epsfxsize=2.0in
    \epsffile{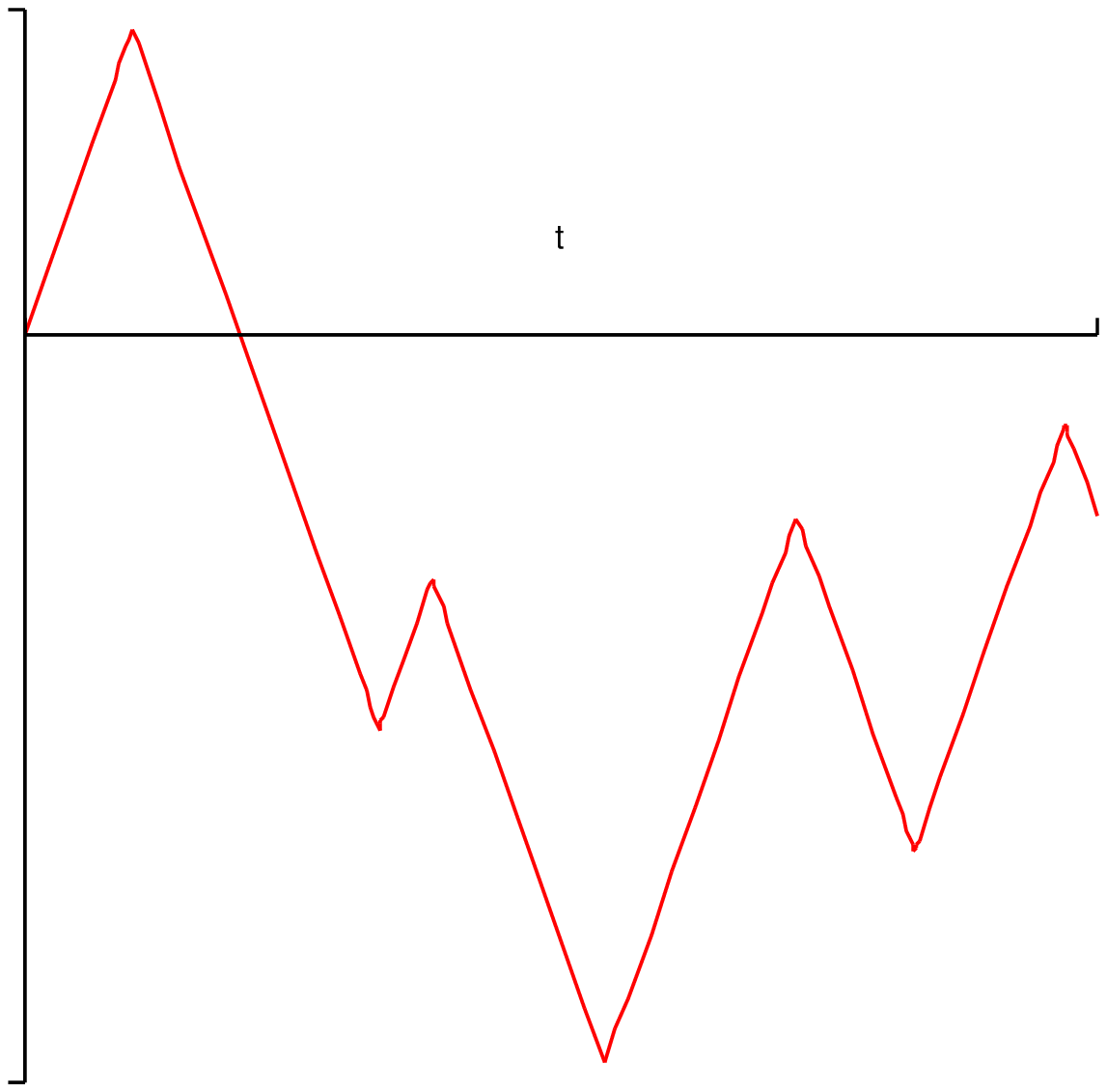}
    }}


\caption{An example $Y(t)$ and $\theta$ }
\label{rtnfig}
\end{figure}

	Two-level telegraph noise, sometimes called popcorn noise or burst noise, appears in a variety of sources. At any given time $t$, the random telegraph signal $Y(t)$, which represents the derivative of the noise, is in either the positive state $\Delta$ or the negative state $-\Delta$. It has probability $\frac{dt}{\tau_1}$ of flipping from the positive state to the negative state, and probability $\frac{dt}{\tau_0}$ of switching from the negative state to the positive state. After time $t$, the parameter of the noise is given by $\theta = \int_0^t Y(t) dt$. If there are no flips, then after time $t$ the parameter will be $\pm \theta_c$, where $\theta_c = \Delta t$. An example of a $Y(t)$ and its corresponding $\theta$ as a function of time is given in Fig. \ref{rtnfig}.

In this paper, we show how to find the expectation value of a function $f(\theta)$, which we shall write as $E[f(\theta)]$. We can write a function in the Fourier basis, and then find the expected values of $E[e^{i m \theta}]$, which we show in Sec. \ref{evendist}, for the case in which the correlation times $\tau_0$ and $\tau_1$ are equal, and in the general case of $\tau_0 \neq \tau_1$ in Sec. \ref{generalrtn}.

In Sec. \ref{control}, we look at how control can be used to suppress effects of this noise on a pair of qubits.
In Sec. \ref{example}, the results are applied to a superconducting qubit system which looks promising for Quantum computing\cite{SVBKWW}.

\section{Evenly distributed Random Telegraph Noise}

\label{evendist}

Here we assume that the correlation time $\tau_c=\tau_0=\tau_1$ is the same in both directions of flips.
Suppose we want to find the expectation value $E[f(\theta)]$, where $\theta$ has a random telegraph distribution. 
\begin{lemma}
\label{telegraphdist}
Since it is Poisson in the number of flips $f$, the distribution of $\theta$ after time $t$ is given by
\begin{equation}
\label{poissondist}
d(\theta) = e^{-\lambda} \sum_{f=0}^{\infty} d(\theta)_f \frac{\lambda^f}{f!}, \text{ for }-\theta_c \leq \theta \leq \theta_c, 
\end{equation}
where the distribution for a given number of flips is on this same domain
\begin{equation}
d_f(\theta) = \begin{cases}
\pm \theta_c & \text{ for $f=0$}\\
\frac{f!}{(\frac{f-1}{2}!)^2} \frac{(\theta_c^2-\theta^2)^\frac{f-1}{2}}{(2\theta_c)^f} & \text{ for odd $f$}\\
d_{f-1}(\theta) & \text{ for even $f > 0 $}
\end{cases}
\end{equation}

\end{lemma}

\begin{proof}
First we look at the discrete case with $q$ steps. $g(b,c) = \binom{b+c-1}{c}$ represents the number of ways to put $c$ identical objects into $b$ different boxes. We have $r$ intervals where we are heading $1$ to the right per unit of time, and $l$ intervals where we are heading $1$ to the left per unit of time. The probability of being at position $j$ afterwards is the coefficient of $x^j$ in 
\begin{equation*}
\sum_{k=0}^q \frac{g(r,k) g(l,q-k)}{g(r+l,q)} x^{2k-q}
\end{equation*}
Since $g(b,c)$ is asymptotically equivalent in $c$ to $\frac{c^{b-1}}{(b-1)!}$, for large $k$ and $q$, this becomes
\begin{align*}
 \frac{(r+l-1)!}{(r-1)!(l-1)!} \sum_{k=0}^q  \frac{k^{r-1} (q-k)^{l-1}}{q^{r+l-1}} x^{2k-q}.
\end{align*}
To get the continuous case, we replace these with $\theta_c=q$, and $\theta=2k-q$, so we have the distribution
\begin{align*}
h_{r,l}(\theta)=\frac{1}{\frac{d\theta}{dk}} \frac{(r+l-1)!}{(r-1)!(l-1)!} \frac{(\frac{\theta_c+\theta}{2})^{r-1}(\frac{\theta_c-\theta}{2})^{l-1}}{\theta_c^{r+l-1}}\\
=\frac{(r+l-1)!}{(r-1)!(l-1)!} \frac{(\theta_c+\theta)^{r-1}(\theta_c-\theta)^{l-1}}{(2\theta_c)^{r+l-1}}.
\end{align*}
Then, if we have $f$ flips, the distribution is
\begin{equation*}
d_f(\theta)=\begin{cases}
h_{\frac{f+1}{2},\frac{f+1}{2}}(\theta)=\frac{f!}{(\frac{f-1}{2}!)^2} \frac{(\theta_c^2-\theta^2)^\frac{f-1}{2}}{(2\theta_c)^f} \text{ for odd $f$}\\
\frac{h_{\frac{f}{2}+1,\frac{f}{2}}(\theta)+h_{\frac{f}{2},\frac{f}{2}+1}(\theta)}{2} = \theta_c \frac{f!}{\frac{f}{2}!(\frac{f}{2}-1)!}\frac{(\theta_c^2-\theta^2)^{\frac{f}{2}-1}}{(2 \theta_c^2)^f}\\ = d_{f-1}(\theta) \text{ for even $f$}.
\end{cases}
\end{equation*}
\end{proof} 

\begin{lemma}
We let $E[g(\theta)]_f$ represent the expected value of an function $g(\theta)$ given that $f$ flips occurred.
If $g(\theta)$ is odd, then $E[g(\theta)]=0$. If $g(\theta)$ is even, then 
\begin{equation}
\label{Egtheta}
E[g(\theta)]_f = \begin{cases}
g(\theta_c)  & \text{ for $f=0$}\\
\frac{f!}{2^{\frac{f-1}{2}} \theta_c^f \frac{f-1}{2}!} g_f(\theta_c) & \text{ for odd $f$}\\
E[g(\theta)]_{f-1} & \text{ for even $f >0$}
\end{cases},
\end{equation}

where 

\begin{align}
g_{-1} (\theta) = \frac{g(\theta)}{\theta} && g_{n+2} (\theta) = \int_{-\theta}^{\theta} \theta g_n (\theta) d \theta.
\end{align}

It then follows from Eq. \ref {poissondist} that the total expected value of an even function $g(\theta)$ can be written as a sum over the $E[g(\theta)]_f$ for odd $f$ as 
\begin{equation}
\label{oddsum}
E[\theta] = e^{-\lambda} \left (g(\theta_c) + \sum_{n=0}^{\infty} E[g(\theta)]_{2n+1} (\frac{\lambda^{2n+1}}{(2n+1)!}+\frac{\lambda^{2n+2}}{(2n+2)!}) \right )
\end{equation}

\end{lemma}

\begin{proof}

If $f$ is odd, then by integration by parts, the expected value of $g(\theta)$ is 
\begin{align*}
E[g(\theta)]_f= \int_{-\theta_c}^{\theta_c} d_f(\theta) g(\theta) d\theta = \frac{f!}{(2\theta_c)^f (\frac{f-1}{2}!)^2}\int_{-\theta_c}^{\theta_c} (\theta_c^2-\theta^2)^{\frac{f-1}{2}} \theta g_{-1}(\theta)  d\theta\\
=(f-1) \frac{f!}{(2 \theta_c)^f (\frac{f-1}{2}!)^2} \int_{-\theta_c}^{\theta_c} (\theta_c^2-\theta^2)^{\frac{f-3}{2}} \theta g_1(\theta) d\theta\\
=(f-1)(f-3)\ldots 2 \frac{f!}{(2\theta_c)^f (\frac{f-1}{2}!)^2} \int_{-\theta_c}^{\theta_c} \theta g_{f-2}(\theta)d\theta\\
= \frac{f!}{2^{\frac{f-1}{2}}\theta_c^f \frac{f-1}{2}!} \frac{g_f(\theta_c)-g_f(-\theta_c)}{2} = f (f-2) \ldots 1 \frac{g_f(\theta_c)-g_f(-\theta_c)}{2 \theta_c^f},
\end{align*}
where $g_{-1}(\theta) = \frac{g(\theta)}{\theta}$, and $g_{n+2}(\theta) = \int \theta g_n(\theta) d\theta$. This is zero for odd functions, and we obtain the desired result for even functions.
\end{proof}

Applying this to $g(\theta)=x^m$ for even $m$ and odd $f$, we get
\begin{equation*}
 E[\theta^m]_f = \Delta^m \frac{1 \times 3 \cdots f}{(m+1)(m+3) \cdots(m+f)} = \Delta^m  \frac{1 \times 3 \cdots (m-1)}{(f+2)(f+4)\cdots (f+m)}.
\end{equation*}
For $m=2$, $E[\theta^2]_f = \frac{\theta_c^2}{f+2}$. Applying this to Eq. \ref{oddsum}, the variance is
\begin{equation}
\label{rtnvar}
\sigma^2 = E[\theta^2]  = \theta_c^2 (\frac{1}{\lambda} + \frac{e^{-2\lambda}-1}{2\lambda^2}) = \Delta^2 t \tau_c + \frac{\Delta^2 \tau_c}{2} (e^{-\frac{2 t}{\tau_c}} -1)
\end{equation}
 The variance is additive with independent noise sources. For $t << \tau_c$, $\sigma^2 \approx \Delta^2 t^2$. For $t >> \tau_c$, $\sigma \approx \Delta^2 (t \tau_c - \frac{1}{2} \tau_c^2)$, and the $\Delta^2 t \tau_c$ part is dominant. 
\begin{theorem}
\label{expectFourier}
$E[\sin (mx)]=0$, and 
\begin{equation}
E[\cos (mx)] = (\frac{1}{2} + \frac{1}{2v}) e^{\lambda (-1 + v)} + (\frac{1}{2} - \frac{1}{2v}) e^{\lambda (-1 - v)},
\end{equation}
where 
\begin{equation*}
v = \sqrt{1-(m \Delta \tau_c)^2}.
\end{equation*}
\end{theorem}

\begin{proof}

For the Fourier transform $e^{i m \theta}$, we get 
\begin{equation}
g_f(\theta) = m^{-\frac{f-1}{2}} \theta_c^{\frac{f+1}{2}}j_{\frac{f-1}{2}}(m\theta),
\end{equation}
where $j_n(x)$ are the Spherical Bessel Functions of the first kind \cite{AS}, which are written as the series
\begin{equation}
j_n(x) = 2^n x^n \sum_{s=0}^{\infty} \frac{(-1)^s (s+n)!}{s! (2s+2n+1)!} x^{2s}.
\end{equation}
Then, the expected value for odd $f$ or even $f+1$ flips is
\begin{equation}
E[e^{im \theta}]_f = \frac{f!}{2^{\frac{f-1}{2}} \frac{f-1}{2}!} \frac{j_{\frac{f-1}{2}}(m \theta_c)}{(m \theta_c)^{\frac{f-1}{2}}} = \frac{f! j_{\frac{f-1}{2}}(z)}{\frac{f-1}{2}! (2z)^{\frac{f-1}{2}}},
\end{equation}
where $z=m \theta_c=mt\Delta$. 
From Eq. \ref{oddsum}, it follows that
\begin{align*}
E[e^{im \theta}]=e^{-\lambda}(\cos(z)+ \sum_{k=0}^{\infty} \frac{(2k+1)! j_k(z)}{k! (2z)^k}(\frac{\lambda^{2k+1}}{(2k+1)!}+\frac{\lambda^{2k+2}}{(2k+2)!}))\\
= e^{-\lambda}(\cos(z) + z \sum_{k=1}^{\infty} \frac{j_{k-1}(z)}{k!}(\frac{\lambda^2}{2z})^k + \lambda \sum_{k=0}^{\infty} \frac{j_k(z)}{k!}(\frac{\lambda^2}{2 z})^k.
\end{align*}
Now since $z j_{-1}(z) = \cos(z)$, and  $j_n(z)$ satisfies the equations
\begin{align*}
\frac{1}{z} \cos \sqrt{z^2 - 2zt} &=& \sum_{n=0}^{\infty} \frac{t^n}{n!} j_{n-1}(z)\\
 \text{sinc} \sqrt{z^2-2zt} &=& \sum_{n=0}^{\infty} \frac{t^n}{n!} j_n(z),
\end{align*}
we have
\begin{align*}
\label{cosexpect}
E[\cos (m \theta)]= E[e^{im \theta}]=e^{-\lambda}(\cos \sqrt{z^2-\lambda^2} + \lambda \text{sinc} \sqrt{z^2-\lambda^2})\\
=e^{-\frac{t}{\tau_c}} \left (\cos \sqrt{(m t \Delta)^2 - (\frac{t}{\tau_c})^2} + \frac{t}{\tau_c} \text{sinc} \sqrt{(m t \Delta)^2 - (\frac{t}{\tau_c})^2} \right)\\
= e^{-\frac{t}{\tau_c}} \left (\cos(t \sqrt{m^2a^2 - \frac{1}{\tau_c^2}}) + \frac{t}{\tau_c} \text{sinc} (t \sqrt{m^2 \Delta^2 - \frac{1}{\tau_c^2}}) \right)\\
= e^{-\frac{t}{\tau_c}} \left (\cosh (\frac{t}{\tau_c}v) + v^{-1} \sinh(\frac{t}{\tau_c}v) \right) = e^{-\lambda} \left (\cosh(\lambda v) + v^{-1} \sinh(\lambda v) \right)
\end{align*}
\end{proof}

\paragraph{Different limits}
For small $\tau_c$ we have $\Delta \tau_c \ll 1$, and therefore
\begin{equation}
\label{likegauss}
E[\cos (m \theta) ] \approx e^{-\frac{1}{2} t m^2 \Delta^2 \tau_c}.
\end{equation}

If $m \Delta \tau_c \gg 1$, then $ E[\cos (m \theta)] \approx e^{-\lambda} \cos (v)$. If in addition, $\tau_c \gg 1$, then $E[\cos (m \theta)] \approx cos(m \theta_c)$.

Up to first order in $\lambda$, we have
\begin{equation}
\label{RTNapprox}
E[\cos (m \theta)] \approx \cos z + \lambda (\text{sinc } z - \cos z) \approx \cos z + \lambda \frac{z^2}{3}.
\end{equation}

\subsection{Multiple sources and Gaussian noise}
\label{gaussian}

If we have multiple independent sources of Random Telegraph Noise, then since $e^{im \theta}$ is a Characteristic Function, the expectation value is the product of the expectation value for each source.
\begin{equation}
E[e^{im \theta}] = e^{-\sum_i \lambda_i} \prod_i \left (\cos \sqrt{(mtx_{0_i})^2-\lambda_i^2} + \lambda_i \text{sinc} \sqrt{(mtx_{0_i})^2-\lambda_i^2} \right)
\end{equation}

If the number of flips $\lambda$ is large, then by the Central limit theorem, the distribution of $\theta$ will approach the Gaussian (also called Normal) distribution 
\begin{equation}
P(x)=\frac{1}{\sigma \sqrt{2 \pi}} e^{-\frac{x^2}{2 \sigma^2}}.
\end{equation}

For a single source of noise, from Eq. \ref{rtnvar}, for large $\lambda$, $\sigma^2 \approx \frac{\theta_c^2}{\lambda} = \Delta^2 \tau_c t$. This is also $(\Delta \tau_c)^2 \lambda$, as one would expect from a random walk. 

If we have $r$ telegraph noise sources, with a mean correlation time of $\tau_{m}$ and a distribution of telegraph strength with mean $\Delta_m$ and standard deviation $\Delta_s$, then $\sigma \approx r(\Delta_m^2 + \Delta_s^2) \tau_m t$. 

\paragraph{$1/f$ noise}
Suppose we have $r$ $1/f$ telegraph noise sources with a frequency in the range $[f_a, f_b]$. The distribution for $f$ is $\frac{1}{f(\log(f_b)-\log(f_a))}$ \cite{Hooge}. Since $\tau_c = \frac{1}{f}$, then $\lambda=ft$, so we have a distribution in $\lambda$ of $\frac{1}{\lambda(\log(\lambda_b)-\log(\lambda_a))}$, where $\lambda_i = f_i t$. From Eq. \ref{rtnvar}, it follows that
\begin{align*}
\sigma^2 = r \int_{\lambda_a}^{\lambda_b} \theta_c^2 (\frac{1}{\lambda} + \frac{e^{-2 \lambda} -1}{2 \lambda^2}) \frac{1}{\lambda(\log(\lambda_b)-\log(\lambda_a))} d\lambda \\ 
= \frac{r \theta_c^2}{\log(\lambda_b)-\log(\lambda_a)} \int_{\lambda_a}^{\lambda_b} (\frac{1}{\lambda^2} + \frac{e^{-2\lambda}-1}{2 \lambda^3}) d\lambda\\
= \frac{r \theta_c^2}{\log(\lambda_b)-\log(\lambda_a)}  \left (\frac{1}{4 \lambda^2} - \frac{1}{\lambda} + \frac{e^{-2 \lambda} (2 \lambda -1)}{4 \lambda^2} - E_1 (-2\lambda) \right) |_{\lambda_a}^{\lambda_b},
\end{align*}
where $E_1(x)$ is the exponential integral. Now if the $\lambda_i$ are large, the $\frac{1}{\lambda}$ part dominates, and so
\begin{equation}
\sigma^2 \approx -\frac{r \theta_c^2}{\lambda(\log(\lambda_b)-\log(\lambda_a))} |_{\lambda_a}^{\lambda_b} = \frac{r t \Delta^2 (\tau_a - \tau_b)}{\log(\tau_a) - \log (\tau_b)}= r t \Delta^2 \frac{\tau_a - \tau_b}{\log(\frac{\tau_a}{\tau_b})},
\end{equation}
where $\tau_i = \frac{1}{f_i}$. This could also be found directly from $\sigma^2 \approx \frac{\theta_c^2}{\lambda}$ for a single source. If $w=\frac{\tau_a}{\tau_b}=\frac{f_b}{f_a}$, then $\sigma^2 = r t \Delta^2 \tau_b \frac{w-1}{\log(w)}$. Since $\tau_a \geq \tau_b$, $r t \Delta^2 \tau_b \leq \sigma^2 \leq r t \Delta^2 \tau_a$. If $r=1$, and $\tau_a = \tau_b = \tau_c$, this gives the single source result.

Sometimes by $1/f$ noise, we mean that the noise has a power spectrum of $\frac{1}{f^{\alpha}}$ for some $\alpha >1 $. In this case, then by a similar calculation to that of \cite{Hooge}, on the domain $[0, \tau_b]$, a power spectrum of $\frac{1}{f^{\alpha}}$ would give a density of random telegraph correlation times of $g(\tau_c) d \tau_c = \frac{(\alpha -1) \tau_c^{\alpha -2}}{\tau_b^{\alpha -1}} d \tau_c$, and so $\tau_m=E[\tau_c] = \int_0^{\tau_b} \tau_c \frac{(\alpha -1) \tau_c^{\alpha -2}}{\tau_b^{\alpha -1}} = \frac{\alpha -1}{\alpha} \tau_b$.

\paragraph{Expected values}
Since $P(x)$ is even, $E[x^n]=0$ for odd $n$. For even $n$, we have
\begin{align*}
E[x^n] = \frac{1}{\sigma \sqrt{2 \pi}} \int_{-\infty}^{\infty} x^n e^{-\frac{x^2}{2 \sigma^2}}\\
 = \frac{1}{\sigma \sqrt{2 \pi}} (-\sigma^2 x^{n-1}e^{-\frac{x^2}{2 \sigma^2}} |_{-\infty}^{\infty} + \int_{-\infty}^{\infty} (n-1) x^{n-2} \sigma^2 e^{-\frac{x^2}{2 \sigma^2}})\\
 = (n-1) \sigma^2 E[x^{n-2}] = \sigma^n (1 \times 3 \cdots (n-1)) = \sigma^n \frac{n!}{2^{\frac{n}{2}} \frac{n}{2}!}.
\end{align*}
Now $E[\sin (mx)] = 0$, and 
\begin{align*}
E[\cos (mx)] = \sum_n \frac{(-m^2)^n E[x^{2n}]}{(2n)!} = \sum_n \frac{(-m^2)^n \sigma^{2n} (2n)!}{(2n)! 2^n n!}
= \sum_n \frac{(\frac{-m^2 \sigma^2}{2})^n}{n!} 
= e^{- \frac{m^2 \sigma^2}{2}}.
\end{align*}
In the case where we have just one flip time $\tau_c$, we get the same result as in Eq. \ref{likegauss}.

\section{Control}
\label{control}
In this section, we assume that the noise is only generated from some Hamiltonian that can be switched on and off. We assume that we need to apply this Hamiltonian for time $t$. 

Lemma \ref{expectFourier} gives the expected values of the Fourier functions without control, assuming that the telegraph starts off in either state with equal probability. Up to second order in $t$, this is
\begin{equation*}
E[\cos m \theta] \approx 1 - \frac{1-v^2}{2} \lambda^2 \approx  1 - \frac{(m \Delta t)^2}{2}.
\end{equation*}
Now, we have the following two possible methods to reduce errors. 

\subsection{Waiting method}
Suppose we wait for a time much greater than the correlation time $t' >> \tau_c$. This will randomize which direction the telegraph is going before we apply a Hamiltonian. If we do this $n$ times, then 
\begin{align*}
E[\cos (m \theta)] = (\frac{1}{2} (1+ v^{-1})e^{\frac{t}{n \tau_c} (-1 + v)} + (1 - v^{-1}) e^{\frac{t}{n \tau_c} (-1-v)})^n\\
 \approx (1 - \frac{m \Delta \frac{t}{n}}{2})^n \approx 1 - \frac{m^2 \Delta^2 t^2}{2 n},
\end{align*}
and $E[\sin (m\theta )]=0$, so by applying the Hamiltonian for time $\frac{t}{n}$ $n$ different times, and waiting a while in between, the rate of error $p$ is changed to $\frac{p}{n}$.

\subsection{Suppressing errors method}
Suppose that we have a Hamiltonian $H$ that we use to create the quantum gate $e^{i t H}$, and it generates a noise Hamiltonian $N$, so that errors are of the form $e^{i \theta N}$ where $\theta$ follows a random telegraph noise distribution. If $H$ and $N$ commute, and we can apply a gate $R$ with a low rate of errors that commutes with $H$ and anti-commutes with $N$, then instead of applying $e^{i t H}$, we apply the gate
\begin{equation*}
(R e^{i \frac{t}{n} H} R^{\dagger} e^{i \frac{t}{n} H})^{\otimes \frac{n}{2}} = e^{i t H}.
\end{equation*}
This breaks the time $t$ into $n$ intervals of equal length, with the direction of the telegraph reversed in between. 
\begin{proposition}
The expected value of $E[\sin (m \theta)]=0$, and
\begin{equation*}
 [\cos (m \theta)] \approx 1 - \frac{m^2 \Delta^2 t^3}{n^2 \tau_c}
\end{equation*}

In this case a rate of error of $p$ is changed to $\frac{2 \lambda}{n^2} p$. 

\end{proposition}
\begin{proof}
Assume that the telegraph starts going off to the right. Then from Eq. \ref{GeneralTelegraph}, since $c=i m$, 
\begin{equation*}
E[e^{i m \theta}] = e^{-\lambda}(\cosh (\lambda v) + (\frac{1}{v} + \frac{im \theta_c}{\lambda v}) \sinh(\lambda v),
\end{equation*}
where $v = \sqrt{1 - (m \Delta \tau_c)^2}$.
Up to third order in $t$, this is
\begin{equation*}
E[e^{i m \theta}]_r \approx (1 - \lambda  + \frac{1}{2} \lambda^2 - \frac{1}{6} \lambda^3) ((1 + \frac{(\lambda v)^2}{2}) + (\frac{1}{v} + \frac{i m t \Delta}{\lambda v})((\lambda v) + \frac{(\lambda v)^3}{6})).
\end{equation*}
Up to second order, this is
\begin{equation*}
E[e^{i m \theta}]_r \approx 1 - \frac{1 - v^2}{2} \lambda^2 + i m t \Delta (1 - \lambda) = 1 - \frac{(m \Delta t)^2}{2} + im t \Delta (1 - \frac{t}{\tau_c})
\end{equation*}
Now suppose that we have two intervals of time $t$ of telegraph noise, with the direction of the telegraph noise reversed in between. The expectation value of $e^{i m \theta}$ for the first is $E[e^{i m \theta}]_r$. Reversing the initial  direction sends $\theta \rightarrow -\theta$ for the second,giving $E[e^{- i m \theta}]_r$, and so the total expected value of $e^{i m \theta}$ is 
\begin{align*}
E[e^{i m \theta}] =  E[e^{i m \theta}]_r E[e^{- i m \theta}]_r \approx (1 - \frac{(m \Delta t)^2}{2})^2 + (m t \Delta (1 - \frac{t}{\tau_c})^2 \\
\approx 1 - 2 m^2 \Delta^2 \tau_c^{-1} t^3 
\end{align*}
For $\frac{n}{2}$ pairs of intervals with $\frac{t}{n}$ time per interval, this becomes
\begin{equation*}
E[e^{i m \theta}] \approx  ( 1 - \frac{2 m^2 \Delta^2 t^3}{n^3 \tau_c})^{\frac{n}{2}}\approx 1 - \frac{m^2 \Delta^2 t^3}{n^2 \tau_c}
\end{equation*}
\end{proof}

From Eq. \ref{GeneralTelegraph}, if the telegraph starts off to the right, instead of $E[\sin (mx)]=0$, we now have
\begin{align*}
E[\sin (mx)] = e^{-\lambda} \frac{m \theta_c}{\sqrt{\lambda^2 - m^2 \theta_c}} sinh (v) \\= e^{-\frac{t}{\tau_c}} \frac{m \Delta}{\sqrt{\tau_c^{-2} - m^2 \Delta^2}} \sinh(t \sqrt{\tau_c^{-2} - m^2 \Delta^2}).
\end{align*} 
Up to first order in $t$, this is $m \Delta t ( 1-\frac{t}{\tau_c})$.

\subsection{Drawbacks}
Both methods assume that the telegraph is equally likely to go in either direction. The waiting method assumes that general decoherence isn't a problem. The suppression method assumes that the gate $R$ can be quickly performed without errors. These assumptions are unrealistic, and other types of noise will likely be created, but these methods could allow for a significant reduction of the magnitude of the  noise.

\section{Example}
\label{example}
From a superconducting qubit system \cite{SVBKWW} we have the noise 
\begin{equation*}
e^{{i \theta Z}^{\otimes 2}}= e^{i \theta (Z \otimes I + I \otimes Z)},
\end{equation*}
where $\theta$ has a random telegraph distribution. Now, 
\begin{equation}
E = \cos^2 \theta I\otimes I + i \sin \theta \cos \theta (I\otimes Z + Z\otimes I) - \sin^2 \theta Z\otimes Z, 
\end{equation}
so the probability of no $Z$ error $n_0=\cos^4 \theta$, each of the probabilities of a $Z$ error on exactly one qubit are $n_1=\sin^2 \theta \cos^2 \theta$, and the probability of a $Z$  error on both qubits is $n_2=\sin^4 \theta$. 
We have that
\begin{multline*}
\vecs
n_0\cr
n_1\cr
n_2\cr
\vece = 
\vecs
E[\cos^4 \theta]\cr
E[\sin^2 \theta \cos^2 \theta]\cr
E[\sin^4 \theta]
\vece =
\mats
\frac{3}{8} & \frac{1}{2} & \frac{1}{8}\cr
\frac{1}{8} & 0 & -\frac{1}{8} \cr
\frac{3}{8} & -\frac{1}{2} & \frac{1}{8} 
\mate 
\vecs
E[1]\cr
E[\cos 2\theta]\cr
E[\cos 4\theta]\cr
\vece ,
\end{multline*}
where the $E[\cos (mx)]$ are given in Thm.. \ref{expectFourier}. 
Then from equation \ref{RTNapprox}, we have that up to 1st order in $\lambda$ and 3nd order in $\theta_c$, 
\begin{align*} 
n_0 &\approx  \cos^4(\theta_c^2) + \frac{4}{3} \lambda \theta_c^2 \\
n_1 &\approx  \sin^2(\theta_c^2) \cos^2(\theta_c^2) - \frac{2}{3} \lambda \theta_c^2 \\
n_2 &\approx \sin^4(\theta_c^2)
\end{align*}

\paragraph{Gaussian noise}
Suppose we assume the noise is Gaussian distributed with a standard deviation of $\sigma$, then 
\begin{align}
n_0 & \approx \frac{3}{8} + \frac{1}{2} e^{-2 \sigma^2} + \frac{1}{8}e^{-8 \sigma^2} &= \frac{3}{8} + \frac{1}{2} r + \frac{1}{8} r^4\\
n_1 & \approx \frac{1}{8} - \frac{1}{8} e^{-8 \sigma^2} & =  \frac{1}{8} - \frac{1}{8} r^4\\
n_2 & \approx \frac{3}{8} - \frac{1}{2} e^{-2 \sigma^2} + \frac{1}{8}e^{-8 \sigma^2} &= \frac{3}{8} - \frac{1}{2} r + \frac{1}{8} r^4,
\end{align}
where $r=e^{-2 \sigma^2}$. 
This gives, up to $\sigma^4$, $n_0 \approx 1 - 2 \sigma^2 + 5 \sigma^4$, $n_1 \approx \sigma^2 - 4 \sigma^4$, $n_2 \approx 3 \sigma^4$. 

\paragraph{Control}
The system \cite{SVBKWW} has a 2 qubit Hamiltonian $H = X \otimes X + Y \otimes Y$, which commutes with the error Hamiltonian of $E_H = I \otimes Z + Z \otimes I$. The gate 
\begin{equation*}
R = X \otimes X
\end{equation*}
commutes with $H$, and anti-commutes with $E_H$. This is very useful, because $R$ is entirely composed of local gates, which have a much lower rate of errors. This is similar to quantum "bang-bang" control \cite{Viola}.

\section{General derivation}
\label{generalrtn}
In section \ref{evendist}, we considered the expectations of a $2$ state random telegraph source with a correlation time $t_c$. In this section we consider that the correlation time depends on which state we're in, that is we have $\frac{dt}{\tau_1}$ chance of flipping per unit time if the telegraph is in the positive state, and $\frac{dt}{\tau_0}$ of flipping time if the telegraph is in the negative state. This is relevant to a physical system at finite temperatures. 

Also we assume that the telegraph starts off in the positive state. Note that if for the positive state $E[e^{im \theta}]_+ = v(m, \tau_0, \tau_1)$, and there is probability of $p_1$ of starting in the positive state, and $p_0$ of starting in the negative state, then 
\begin{equation*}
E[e^{im \theta)}] = p_0 E[e^{im (-\theta)}]_- + p_1 E[e^{im \theta}]_+ = p_0 v(-m, \tau_1, \tau_0) + p_1 v(m, \tau_0, \tau_1).
\end{equation*}

\begin{lemma}
\label{distlemma}
Suppose the telegraph starts off in the positive state (so that the noise parameter is increasing). 
The distribution of $\theta$ is given by a sum over all of the possible number of flips 
\begin{equation*}
d(\theta) = \sum_{f=0}^{\infty} d_f(\theta), 
\end{equation*}
where 
\begin{align}
d_f(\theta) = \begin{cases} 
h(\theta) \delta (\theta - \theta_c) & \text{ if } f=0 \cr
h(\theta) \frac{(\lambda_0 \lambda_1)^{\frac{f}{2}}}{(2 \theta_c)^f \frac{f}{2}! (\frac{f}{2}-1)!} (\theta_c + \theta) (\theta_c^2 - \theta^2)^{\frac{f}{2}-1}                            & \text{ for even } f> 0 \cr
h(\theta) \frac{\lambda_1 (\lambda_0 \lambda_1)^{\frac{f-1}{2}}}{(2\theta_c)^f (\frac{f-1}{2}!)^2}  (\theta_c^2 - \theta^2)^{\frac{f-1}{2}}  & \text{ for odd } f \cr
\end{cases}\\
h(\theta) = e^{\lambda_0 \frac{\theta - \theta_c}{2 \theta_c}} e^{-\lambda_1 \frac{\theta + \theta_c}{2 \theta_c}} 
\end{align}

on the domain $[-\theta_c,\theta_c]$, where $\theta_c = t \Delta$, $\lambda_i = \frac{t}{\tau_i}$.
\end{lemma}
\begin{proof}
 
There are no flips with probability $e^{-\lambda_1}$, in which case $\theta = \theta_c$, which gives $d_0(\theta)=e^{-\lambda_1} \delta (\theta-\theta_c)$.

Suppose we have $r$ intervals where the telegraph is in the positive state and so the noise parameter is increasing, and $l$ intervals where the telegraph noise is in the negative state and so the noise parameter is decreasing, and we have $q$ steps. Then we have $g(r,k) g(l, q-k)$ ways to end up at position $2k-q$. Suppose we start off in the increasing telegraph state, and the probability of flipping if we're increasing is $\tau_1^{-1}$, and if we're decreasing is $\tau_0^{-1}$. The probability of being at position $j$ afterwards is the coefficient of $x^j$ in
\begin{align*}
\sum_{k=0}^q (1-\tau_1^{-1})^{k-l} (\tau_1^{-1})^l (1-\tau_0)^{q-k-r+1} (\tau_0^{-1})^{r-1} g(r,k) g(l,q-k) x^{2k-q}\\
\approx e^{-\frac{k}{\tau_1}} (\tau_1^{-1})^l e^{-\frac{q-k}{\tau_0}} (\tau_0^{-1})^{r-1} \frac{k^{r-1} (q-k)^{l-1}}{(r-1)!(l-1)!} x^{2k-q}
\end{align*}
We make this continuous, so we have $\theta = 2k-q$, $\theta_c=q$, and pick up a factor $\frac{1}{2}$ from $\frac{d k}{d \theta}$, and so have 

\begin{align*}
\frac{1}{2} e^{-\frac{\theta_c +\theta}{2 \tau_1}} \tau_1^{-l} e^{-\frac{\theta_c-\theta}{2 \tau_0}} \tau_0^{1-r} \frac{(\frac{\theta_c+\theta}{2})^{r-1} (\frac{\theta_c - \theta}{2})^{l-1}}{(r-1)! (l-1)!} \\
= h(\theta)   \frac{(\theta_c+\theta)^{r-1} (\theta_c - \theta)^{l-1}}{2^{r+l-1} \tau_1^l \tau_0^{r-1} (r-1)! (l-1)!}
\end{align*}

If we have an even number of telegraph flips $f > 0$, then $r=\frac{f}{2}+1$, $l = \frac{f}{2}$. If we have an odd number of telegraph flips, then $r=l=\frac{f+1}{2}$. Plugging these into the previous equation produces the desired result.
\end{proof}

\begin{lemma}
\label{intlemma}

\begin{align*}
a_n =  \int_{-\theta_c}^{\theta_c}  e^{c \theta} \theta(\theta_c^2 - \theta^2)^n d \theta = \frac{2^n n! (-1)^n}{c^{2n+3} \theta_c} (p_{n+2}(-c \theta_c)e^{c \theta_c} + p_{n+2}(c \theta_c) e^{-c\theta_c}),
\end{align*}
where the $p_n(x)$ are the Carlitz Bessel polynomials described in \cite{BP}. 
\end{lemma}

\begin{proof}

If we let 
\begin{align*}
g_0(\theta) = e^{c \theta}\\
g_{n+1}(\theta) = \int g_n(\theta) \theta d \theta,
\end{align*}
then by integration by parts, 
\begin{align*}
\int_{-\theta_c}^{\theta_c} \theta g_k(\theta) (\theta_c^2 - \theta^2)^n d \theta\\ 
= g_{k+1}(\theta) (\theta_c^2 - \theta^2)^n |_{-\theta_c}^{\theta_c} - \int_{-\theta_c}^{\theta_c} g_{k+1}- 2 n \theta (\theta_c^2 - \theta^2)^{n-1} d\theta \\
= 2n \int_{-\theta_c}^{\theta_c}  g_{k+1}- 2 n \theta (\theta_c^2 - \theta^2)^{n-1} d\theta 
= 2^n n! (g_{n+k+1}(\theta_c) - g_{n+k+1}(-\theta_c)) , 
\end{align*}
so 
\begin{equation*}
a_n = 2^n n! (g_{n+1}(\theta_c) - g_{n+1}(-\theta_c)).
\end{equation*}
Now, from the the formula for the Carlitz Bessel functions, it can be shown that they satisfy the differential equation
\begin{equation*}
p'_{n+1}(x) - p_{n+1}(x)(1 + x^{-1}) + x p_n(x) = 0. 
\end{equation*}
Note that this implies that $p_0(x) = 1$.
If we let 
\begin{equation*}
g_n (\theta) = - \frac{p_{n+1}(-c \theta)}{c \theta} \frac{(-1)^n}{c^{2n}} e^{c \theta},
\end{equation*}
then this satisfies the differential equations $g'_{n+1}(\theta) = \theta g_n(\theta)$, $g_0(\theta)=e^{c \theta}$, and so we get the desired result. 

\end{proof}

\begin{theorem}

The expectation of a characteristic function $e^{i m \theta}$ is
\begin{equation}
\label{GeneralTelegraph}
E[e^{i m \theta}] = e^{-\frac{\lambda_0 + \lambda_1}{2}} (\cosh u + \frac{\lambda_1 + c \theta_c}{u} \sinh u)
\end{equation}
where 
\begin{equation}
u = \sqrt{c^2 \theta_c^2 + \lambda_0 \lambda_1} = t \sqrt{c^2 \Delta^2 + \tau_0^{-1} \tau_1^{-1}},
\end{equation}
where $c = \frac{\lambda_0 - \lambda_1}{2 \theta_c} + i m$.

\end{theorem}
\begin{proof}
From Lemma \ref{distlemma}, the contribution from an odd number of flips $f$ is
\begin{align*}
E[e^{i m \theta}]_{\text{odd}} = \sum_{n=0}^{\infty} \int_{-\theta_c}^{\theta_c} e^{i m \theta} d_{2n+1}(\theta) d \theta \\
= e^{-\frac{\lambda_0 + \lambda_1}{2}} \lambda_1 \sum_{n=0}^{\infty} \frac{(\lambda_0 \lambda_1)^n}{(2 \theta_c)^{2n+1} (n!)^2} \int_{-\theta_c}^{\theta_c} (\theta_c^2 - \theta^2)^n e^c d \theta,
\end{align*}

By integration by parts,
\begin{equation*} 
\int_{-\theta_c}^{\theta_c}  e^{c \theta} (\theta_c^2 - \theta^2)^n d\theta = \frac{2n}{c} a_{n-1},
\end{equation*}
and so from Lemma \ref{intlemma}, if $z=c \theta_c$, 
\begin{align*}
E[e^{i m \theta}]_{\text{odd}} = - \lambda_1 \frac{e^{- \frac{\lambda_0+\lambda_1}{2}}}{2 z^2} \sum_n (-\frac{\lambda_0 \lambda_1}{2 z^2})^n \frac{1}{n!} (p_{n+1}(-z) e^z + p_{n+1}(z) e^{-z})
\end{align*}
Now, by differentiating the formula 
\begin{equation*}
\sum_{k=0}^{\infty} \frac{p_k(x)}{k!} t^k = e^{x (1 - \sqrt{1 - 2 t})}
\end{equation*}
from \cite{BP}, we get
\begin{equation*}
\sum_{k=0}^{\infty} \frac{p_{k+1}(x)}{k!} t^k = \frac{x e^{x (1 - \sqrt{1-2t})}}{\sqrt{1 - 2t}},
\end{equation*}
so if $t= - \frac{\lambda_0 \lambda_1}{2 z^2}$, 
\begin{align*}
  E[e^{i m \theta}]_{\text{odd}} =  - \lambda_1 e^{-\frac{\lambda_0 + \lambda_1}{2}} \frac{-z e^{-z(1-\sqrt{1-2t}} e^z +z e^{z (1-\sqrt{1-2t})} e^{-z}}{2 z^2 \sqrt{1-2t}}\\
= \lambda_1 e^{-\frac{\lambda_0 + \lambda_1}{2}} \frac{\sinh (z \sqrt{1-2t})}{z \sqrt{1-2t}} = 
\lambda_1 e^{\frac{\lambda_0 + \lambda_1}{2}} \frac{\sinh u}{u}
\end{align*}

For even $f>0$, the contribution is
\begin{equation*}
E[e^{i m \theta}]_{\text{even }>0} = e^{-\frac{\lambda_0 + \lambda_1}{2}} \sum_{n=1}^{\infty} \frac{(\lambda_0 \lambda_1)^n}{(2 \theta_c)^{2n} n! (n-1)!} \int_{-\theta_c}^{\theta_c} (\theta_c^2 - \theta^2)^{n-1} (\theta_c + \theta) e^{c \theta} d \theta,
\end{equation*}
and 
\begin{equation*}
\int_{-\theta_c}^{\theta_c} (\theta_c^2 - \theta^2)^{n-1} (\theta_c + \theta) e^{c \theta} d \theta = \frac{2 (n-1) \theta_c}{c} a_{n-2} + a_{n-1}, 
\end{equation*}
 so if $z=c \theta_c$, 
\begin{align*}
E[e^{i m \theta}]_{\text{even }>0} = \frac{e^{-\frac{\lambda_0 + \lambda_1}{2}}}{2} \sum_{n=1}^{\infty}  (-\frac{\lambda_0 \lambda_1}{2 z^2})^n \frac{1}{n!}\\
((p_n(-z)e^z + p_n(z)e^{-z})-   \frac{p_{n+1}(-z)e^{z} + p_{n+1}(z)e^{-z}}{z}).
\end{align*}
In the sum, if $n=0$, 
\begin{align*}
 ((p_0(-z)e^z + p_0(z)e^{-z}-\frac{p_1(-z)e^z + p_1(z)e^{-z}}{z})\\
= \frac{e^{-\frac{\lambda_0 + \lambda_1}{2}}}{2} ((e^z + e^{-z}) - \frac{-z e^z + z e^{-z}}{z}) = e^{z - \frac{\lambda_0 + \lambda_1}{2}} = e^{-\lambda_1} e^{i m \theta_c},
\end{align*}
so we can include $f=0$, and so we get a sum over all even $f$, with $t=-\frac{\lambda_0 \lambda_1}{2 z^2}$,
\begin{align*}
E[e^{i m \theta}]_{\text{even }} = \frac{e^{-\frac{\lambda_0 + \lambda_1}{2}}}{2} \sum_{n=0}^{\infty} \frac{t^n}{n!} (p_n(-z)e^z + p_n(z)e^{-z})\\
 - \frac{e^{-\frac{\lambda_0 + \lambda_1}{2}}}{2} \sum_{n=0}^{\infty} \frac{t^n}{n!} (p_{n+1}(-z)e^z + p_{n+1}(z)e^{-z})\\
= \frac{e^{-\frac{\lambda_0 + \lambda_1}{2}}}{2} (e^{-z(1-\sqrt{1-2t})} e^z + e^{z(1-\sqrt{1-2t})} e^{-z}) + z e^{\frac{\lambda_0 + \lambda_1}{2}} \frac{\sinh u}{u}\\
= e^{\frac{\lambda_0 + \lambda_1}{2}} (\cosh u + \frac{z}{u} \sinh u)
\end{align*}

\end{proof}

\section{Acknowledgments}
We thank the NSF for financial support under ITR Grant No. EIA-0205641. 

We thank K. B. Whaley for helpful discussions.

\end{document}